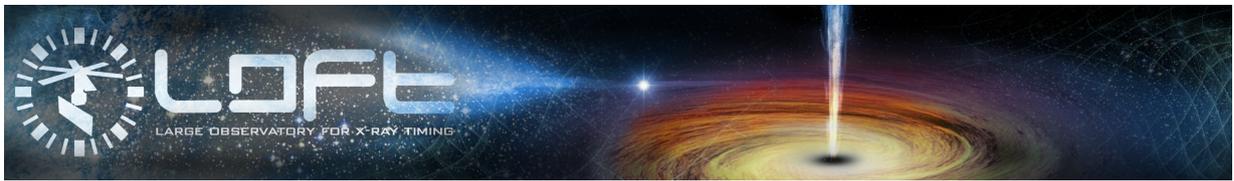

# The *LOFT* perspective on neutron star thermonuclear bursts

## White Paper in Support of the Mission Concept of the Large Observatory for X-ray Timing

### Authors

J.J.M. in 't Zand[1], D. Altamirano[2], D.R. Ballantyne[3], S. Bhattacharyya[4], E.F. Brown[5], Y. Cavecchi[6], D. Chakrabarty[7], J. Chenevez[8], A. Cumming[9], N. Degenaar[10], M. Falanga[11], D.K. Galloway[12], A. Heger[12], J. José[13], L. Keek[3], S. Linares[14], S. Mahmoodifar[15], C.M. Malone[16], M. Méndez[17], M.C. Miller[18], F.B.S. Paerels[19], J. Poutanen[20], A. Różańska[21], H. Schatz[22], M. Serino[23], T.E. Strohmayer[24], V.F. Suleimanov[25], F.-K. Thielemann[26], A.L. Watts[6], N.N. Weinberg[7], S.E. Woosley[27], W. Yu[28], S. Zhang[29], M. Zingale[30]

[1] SRON Netherlands Institute for Space Research, Sorbonnelaan 2, 3584 CA Utrecht, the Netherlands
[2] School of Physics & Astronomy, University of Southampton, Highfield, Southampton, SO17 1BJ, UK
[3] Center for Relativistic Astrophysics, School of Physics, Georgia Institute of Technology, Atlanta, GA 30332, USA
[4] Department of Astronomy and Astrophysics, Tata Institute of Fundamental Research, 1 Homi Bhabha Road, Mumbai 400005, India
[5] Department of Physics and Astronomy, National Superconducting Cyclotron Laboratory and the Jopint Institute for Nuclear Astrophysics, Michigan State University, East Lansing, MI 48824, USA
[6] Anton Pannekoek Institute, University of Amsterdam, Science Park 904, 1098 XH Amsterdam, The Netherlands
[7] Dept. Physics and Kavli Institute for Astrophysics and Space Research, Massachusetts Institute of Technology, Cambridge, MA 02139, USA
[8] National Space Institute, Technical University of Denmark, Electrovej 327, 2800 Lyngby, Denmark
[9] Physics Dept., McGill University, 3600 Rue University, Montreal, QC, H3A 2T8, Canada
[10] Institute of Astronomy, University of Cambridge, Madingley Road, Cambridge CB3 OHA, UK
[11] International Space Science Institute, Hallerstrasse 6, 3012 Bern, Switzerland
[12] Monash Centre for Astrophysics, School of Physics & Astrophysics, Monash University, VIC 3800, Australia
[13] Departament de Física i Enginyeria Nuclear, EUETIB, Universitat Politècnica de Catalunya, 08036 Barcelona, Spain; and Institut d'Estudis Espacials de Catalunya, 08034 Barcelona, Spain
[14] Instituto de Astrofísica de Canarias, c/ Vía Láctea s/n, 38205 La Laguna, Tenerife, Spain; and Universidad de la Laguna, Departamento de Astrofísica, 38206 La Laguna, Tenerife, Spain
[15] CRESST and X-ray Astrophysics Laboratory, NASA/GSFC, Greenbelt, MD 20771, USA; Department of Astronomy, University of Maryland College Park, MD 20742, USA
[16] Los Alamos National Laboratory, CCS-2, Los Alamos, NM 87545, USA
[17] Kapteyn Astronomical Institute, University of Groningen, P.O. Box 800, 9700 AV Groningen, the Netherlands
[18] Department of Astronomy and Joint Space-Science Institute, University of Maryland, College Park, MD 20742-2421, USA
[19] Columbia Astrophysics Laboratory, 550 West 120th Street, New York, NY 10027, USA
[20] Tuorla Observatory, Department of Physics and Astronomy, University of Turku, Väisäläntie 20, 21500 Piikkiö, Finland
[21] N. Copernicus Astronomical Center PAS, Bartycka 18, 00-716 Warsaw, Poland
[22] National Superconducting Cyclotron Laboratory, Michigan State University, 640 S. Shaw Lane, East Lansing, Michigan 48824, USA
[23] Institute of Physical and Chemical Research (RIKEN), 2-1 Hirosawa, Wako, Saitama 351-0198, Japan
[24] Astrophysics Science Division and Joint Space-Science Institute, NASA's Goddard Space Flight Center, Greenbelt, MD 20771, USA
[25] Institut für Astronomie und Astrophysik, Kepler Center for Astro and Particle Physics, Universität Tübingen, Sand 1, 72076 Tübingen, Germany
[26] Departement Physik, Universität Basel, 4056 Basel, Switzerland
[27] Department of Astronomy and Astrophysics, University of California Santa Cruz, CA 95064, USA
[28] Shanghai Astronomical Observatory, Chinese Academy of Sciences, 80 Nandan Road, Shanghai 200030, China
[29] Laboratory for Particle Astrophysics, Institute of High-Energy Physics, Beijing 100049, China
[30] Department of Physics & Astronomy, Stony Brook University, Stony Brook, NY 11794-3800, USA

## Preamble

The Large Observatory for X-ray Timing, *LOFT*, is designed to perform fast X-ray timing and spectroscopy with uniquely large throughput (Feroci et al. 2014). *LOFT* focuses on two fundamental questions of ESA's Cosmic Vision Theme "Matter under extreme conditions": what is the equation of state of ultra-dense matter in neutron stars? Does matter orbiting close to the event horizon follow the predictions of general relativity? These goals are elaborated in the mission Yellow Book (`http://sci.esa.int/loft/53447-loft-yellow-book/`) describing the *LOFT* mission as proposed in M3, which closely resembles the *LOFT* mission now being proposed for M4.

The extensive assessment study of *LOFT* as ESA's M3 candidate mission demonstrates the high level of maturity and the technical feasibility of the mission, as well as the scientific importance of its unique core science goals. For this reason, the *LOFT* development has been continued, aiming at the new M4 launch opportunity, for which the M3 science goals have been confirmed. The unprecedentedly large effective area, large grasp, and spectroscopic capabilities of *LOFT*'s instruments make the mission capable of state-of-the-art science not only for its core science case, but also for many other open questions in astrophysics.

*LOFT*'s primary instrument is the Large Area Detector (LAD), a $8.5\,\mathrm{m}^2$ instrument operating in the 2–30 keV energy range, which will revolutionise studies of Galactic and extragalactic X-ray sources down to their fundamental time scales. The mission also features a Wide Field Monitor (WFM), which in the 2–50 keV range simultaneously observes more than a third of the sky at any time, detecting objects down to mCrab fluxes and providing data with excellent timing and spectral resolution. Additionally, the mission is equipped with an on-board alert system for the detection and rapid broadcasting to the ground of celestial bright and fast outbursts of X-rays (particularly, Gamma-ray Bursts).

This paper is one of twelve White Papers that illustrate the unique potential of *LOFT* as an X-ray observatory in a variety of astrophysical fields in addition to the core science.

# 1 Summary

The Large Area Detector (LAD) on the Large Observatory For X-ray Timing (*LOFT*), with a 8.5 m$^2$ photon-collecting area in the 2–30 keV bandpass at CCD-class spectral resolving power ($\lambda/\Delta\lambda = 10$–100), is designed for optimum performance on bright X-ray sources. Thus, it is well-suited to study thermonuclear X-ray bursts from Galactic neutron stars. These bursts will typically yield $2 \times 10^5$ photon detections per second in the LAD, which is at least 15 times more than with any other instrument past, current or anticipated. The Wide Field Monitor (WFM) foreseen for *LOFT* uniquely combines 2–50 keV imaging with large (30%) prompt sky coverage. This will enable the detection of tens of thousands of thermonuclear X-ray bursts during a 3-yr mission, including tens of superbursts. Both numbers are similar or more than the current database gathered in 50 years of X-ray astronomy.

*LOFT* will enable great advances in the understanding of thermonuclear X-ray bursts, in particular on the following questions.

- *What is the nature of burst oscillations?* Oscillations in burst tails can be studied at much better sensitivity. This can increase the percentage of bursts with detected oscillations, enable the detection of shorter oscillation trains and faster frequency drifts, probe fainter amplitudes and make possible spectro-timing analyses. Thus, *LOFT* will tremendously enlarge our knowledge about burst oscillations including the possible persistence of hot spots and the potential roles of surface $r$ and $g$ modes.

- *Is there a preferred ignition latitude on the neutron star (as expected for fast spinning neutron stars with anisotropic accretion); how do Coriolis, thermal and hydrodynamic effects combine to control flame spread?* Thermonuclear flame spreading can be studied with the LAD at an unprecedented level of detail, through burst oscillations during the rise of the burst. Modeling the evolution of the spectrum and amplitude will constrain the flame spreading physics.

- *What is the composition of the ashes of nuclear burning; how large is the gravitational redshift on the neutron star surface?* The LAD will be uniquely sensitive to details of the continuum spectrum that are induced by Compton scattering in the atmosphere and to absorption features from dredged-up nuclear ashes such as iron and nickel which may enable direct compositional studies. Surface absorption features will be gravitationally redshifted, which will provide additional constraints on the compactness of neutron stars.

- *What burns in superbursts; what determines the thermodynamic equilibrium of the crust and ocean; what determines the stability of nuclear burning; what is the nature of surface nuclear burning at low accretion rates?* WFM observations will measure precisely the superburst recurrence time and time profile. This will constrain the thermodynamic balance in the ocean and crust and probe in unprecedented detail the boundary between stable and unstable thermonuclear burning. WFM will accumulate large exposure times on all bursters (including yet unknown ones) and, thus, detect rare classes of events, such as the most powerful intermediate-duration bursts at very low accretion rates.

- *How does the accretion emission change during bursts?* *LOFT*'s unprecedented sensitivity at photon energies beyond 20 keV, will reduce an important source of uncertainty in burst studies.

# 2 Introduction

Thermonuclear X-ray bursts are due to thermonuclear shell flashes in the matter accreted onto neutron stars from the atmosphere of a Roche-lobe overfilling companion star in a low-mass X-ray binary (LMXB; for reviews see Lewin et al. 1993; Strohmayer & Bildsten 2006; Galloway et al. 2008). The ignition conditions are set by the ambient temperature, mass accretion rate, fuel composition and neutron star spin and magnetic field. The ambient temperature is set by stable nuclear processes in the ocean and crust, whose yields are a direct function of the mass accretion rate (on short as well as long time scales). Ignition starts at a certain location and depth in the neutron star ocean and spreads over the complete ocean in less than a few seconds. At each location, most energy is released in a fraction of a second. The radiative flux can be larger than the Eddington limit, resulting in photospheric expansion and potential mass loss. The subsequent cooling of the burnt layer is what one observes as an X-ray burst with a thermal spectrum with temperatures that peak between 2 and 3 keV. The cooling lasts from 10 to $10^5$ s, for ignition column depths from $10^8$ to $10^{12}$ g cm$^{-2}$. The constituent mostly responsible for the runaway process is helium, except for the deepest ignition where it is presumably carbon (like in type Ia supernovae). With sufficiently large instrument sensitivity, oscillations are often detected in bursts with a frequency close to the neutron star spin frequency (see reviews by Galloway et al. 2008; Watts 2012).

While X-ray bursts are well understood as a phenomenon associated with thermonuclear burning, many important questions are unanswered. Some of those are fundamental, being related to the structure of the neutron star and exotic nuclear processes. The Large Observatory for X-ray Timing (*LOFT*) will provide the opportunity to answer these questions. With the anticipated observation program, it is expected that the Large Area Detector (LAD) will detect at least several hundreds of X-ray bursts, with about 15 times as much effective area (i.e., 8.5 m$^2$) and 5 times finer spectral resolution (i.e., 260 eV) as the previous burst workhorse RXTE-PCA (Jahoda et al. 2006) and the forthcoming Astrosat-LAXPC (Paul & LAXPC Team 2009). A typical peak photon count rate will be 2×10$^5$ s$^{-1}$. The coded-mask imaging Wide Field Monitor (WFM), thanks to its 4 sr field of view (30% of the sky, which is six times larger than RXTE-ASM and four times larger than BeppoSAX-WFC) will detect as many as ten to twenty thousand X-ray bursts over a 3-yr mission, which is similar to how many were detected throughout the present history of X-ray astronomy. It is clear that *LOFT* will provide a rich and novel data set on X-ray bursts. While the primary science motives for X-ray burst observations with *LOFT* are dense matter physics and gravity in the strong field regime (Feroci et al. 2014), the secondary motives discussed here concern their fundamental understanding, in connection with nucleosynthesis, hydrodynamics, flame spreading and accretion flows. As a benefit, the outcome of these studies will help reduce the risk of any systematic errors in studies for *LOFT*'s primary science goals. This White Paper briefly discusses interesting questions in four areas: nuclear reactions (§ 3), non-uniform emission patterns (§ 4), atmospheric conditions (§ 5) and accretion flows (§ 6).

## 3   Nuclear reactions

The primary nuclear processes active in thermonuclear burning on NSs are the CNO cycle to burn hydrogen ($\beta$-limited CNO cycle above $\sim 8 \times 10^7$ K), the triple-$\alpha$ process to burn helium (above several $10^8$ K), the $\alpha$p-process above $5 \times 10^8$ K to produce elements like Ne, Na and Mg, and the rapid proton capture process ('rp process', above $10^9$ K) to burn hydrogen into even heavier elements (e.g., Fujimoto et al. 1981; Wallace & Woosley 1981; Fisker et al. 2008). In cases of hydrogen-free helium accretion or where the $\beta$-limited CNO cycle burns all hydrogen to helium before ignition, bursts are due to ignition in the helium layer. If beta-limited H burning is not proceeding continuously, the CNO cycle may eventually ignite in the hydrogen layer, taking helium ignition along. The ignition of helium may be delayed in that regime (Peng et al. 2007; Cooper & Narayan 2007a), resulting in pure hydrogen flashes.

Hundreds of proton-rich isotopes are produced during thermonuclear burning, particularly by the rp process, yielding rare isotopes that have not been made in laboratories yet.

### 3.1   Superbursts as probes of the neutron star crust

In the past one and a half decades, two new, long and rare kinds of X-ray bursts have been discovered: 'superbursts' (Cornelisse et al. 2000; Kuulkers et al. 2002b; Strohmayer & Brown 2002) and 'intermediate duration bursts' (in 't Zand et al. 2005; Cumming et al. 2006) which ignite at $10^2$ to $10^4$ larger column depths than ordinary bursts. It has been proposed that superbursts are fueled by carbon (Cumming & Bildsten 2001; Strohmayer & Brown 2002), but it is unclear how the carbon can survive the rp-process or ignite (Schatz et al. 2003; in 't Zand et al. 2003; Keek et al. 2008) and whether superbursts are sometimes not fueled by carbon (Kuulkers et al. 2010). In contrast, it seems clear that intermediate duration bursts are fueled by helium on relatively cold NSs.

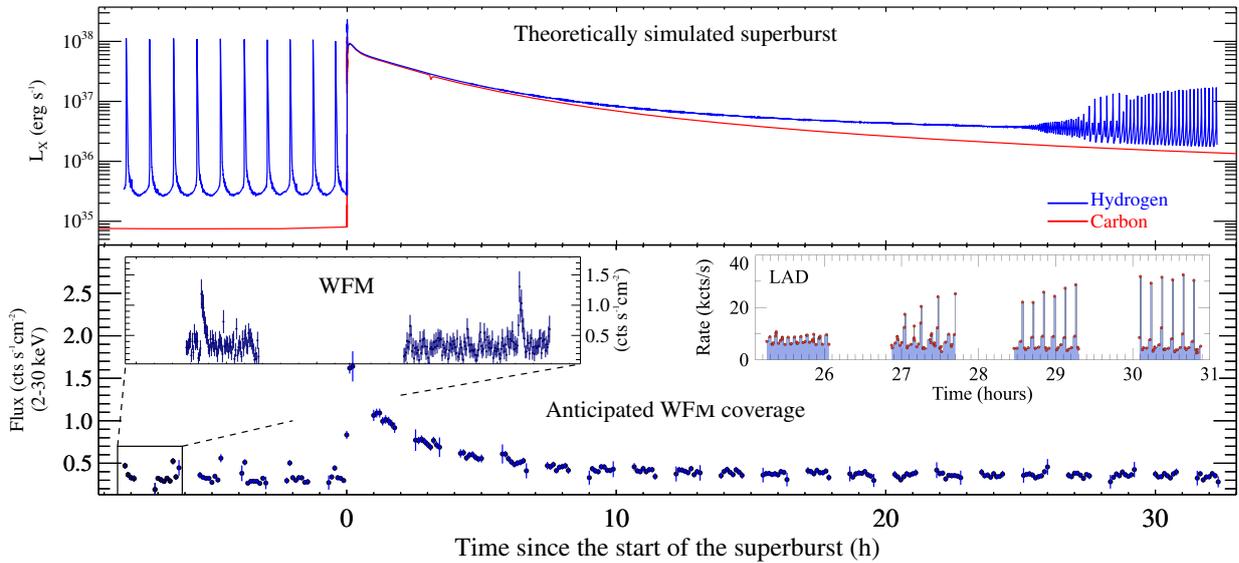

**Figure 1:** Simulation of the ordinary bursting behavior (blue curve) around a superburst (red curve) in a hydrogen-rich LMXB (from Keek et al. 2012). The superburst quenches ordinary bursts for one day after which a regime of marginal burning commences, characterized by mHz QPOs. The lower panel shows possible WFM and LAD measurements of such a superburst, if it is at 8 kpc, on-axis, placed at 45° from the Galactic center and if the accretion rate is 10% of the Eddington limit.

Due to their larger ignition depth, these long bursts can serve as probes of the thermal properties of the crust, one of the outstanding questions of neutron star research. Superbursts have a property which is intriguing in light of this dependence: they ignite sooner than expected on the basis of the measured mass accretion rate and presumed crust properties (Keek et al. 2008). It looks as though the thermal balance in the crust (determined by nuclear reactions in the crust and neutrino cooling in the core) is different from expected (see also Schatz et al. 2014). Freeze out of light elements at the base of the ocean may play a role in the thermal balance of the ocean (Medin & Cumming 2011). Better measurements are needed to constrain more accurately the recurrence time and the time history of the accretion between superbursts (Fig. 1). The high duty cycle of *LOFT*-WFM (e.g., 10 Msec of exposure per year on the Galactic Center field) will be instrumental in this, and should enable the detection of tens of superbursts of different sources in just a few years. High duty cycle observations can also help to expand the database on superburst onsets and peak luminosities, which may help to constrain the fuel mixture (i.e., the carbon abundance; Cumming & Macbeth 2004).

## 3.2 Unstable versus stable nuclear burning

A long standing issue in X-ray burst research is the threshold between stable and unstable helium burning (van Paradijs et al. 1988; Cornelisse et al. 2003a; Zamfir et al. 2014). Thermonuclear burning becomes stable once the temperature becomes so high that the temperature dependence of the energy generation rate levels off to that of the radiative cooling rate (which is proportional to $T^4$). More than 99% of all X-ray bursts are due to ignition of helium through $3\alpha$ burning and $\alpha$-captures. It is predicted that the ambient temperatures in the burning layer only become high enough ($\gtrsim 5 \times 10^8$ K) for stable burning when the mass accretion rate is above the Eddington limit (e.g., Bildsten 1998). However, one finds evidence of stable burning (i.e., absence of X-ray bursts and presence of mHz QPOs indicative of marginally stable burning) already at 10–50% of Eddington (e.g., Revnivtsev et al. 2001; Cornelisse et al. 2003b; Heger et al. 2007b; Altamirano et al. 2008; Linares et al. 2012). A different perspective on this issue may be provided by *LOFT*, by following extensively the aftermath of superbursts for about a week (see above). The superburst is another means to increase the ambient temperatures. While the NS ocean cools down in the hours to days following a superburst, models predict (Cumming & Macbeth 2004; Keek et al. 2012, see Fig. 1) that it is possible to follow the transition from stable to unstable burning on a convenient time scale, through sensitive measurements of low-amplitude ($\sim 0.1\%$) mHz oscillations and the return of normal, initially faint, bursting activity.

### 3.3 Pure hydrogen flashes

Pure hydrogen flashes are expected to occur in a narrow range of mass accretion rates somewhere between roughly 0.1 and 1% of the Eddington limit with typical peak luminosities $10^{-3}$ times that of helium-fueled bursts and recurrence times of order 1 day (Peng et al. 2007; Cooper & Narayan 2007a). They provide a path towards deep ignition of helium in hydrogen-rich accretion conditions, should they occur. Pure hydrogen flashes have not been detected yet, because no instrument has been sensitive enough to do that. That will change with *LOFT*. The peak flux is anticipated to be of order 1–10 mCrab, implying a LAD photon count rate of order 200–2000 s$^{-1}$ with a black body like spectrum of temperature $kT \approx 0.5$ keV which should be easily detectable. If they were detected, these flashes would provide the possibility to probe sedimentation of He and CNO out of the ignition layer and of the thermal state of crust and ocean at low accretion rates. This finding would have important implications for accretion physics as it would show that accretion disks can remain hot down to lower accretion rates than accretion disk instability theory indicates (e.g., Lasota 2001). Furthermore, it would imply that intermediate duration bursts from systems with low accretion rates do not necessarily point to hydrogen-deficient accretion such as in ultracompact X-ray binaries (e.g., in 't Zand et al. 2007).

### 3.4 WFM measurements on nuclear burning

The WFM is expected to detect possibly as many as 20 thousand X-ray bursts, which is more than have been detected throughout the whole era of X-ray astronomy so far, and an order of magnitude more than for any individual instrument (for instance, the wide-field camera BeppoSAX-WFC detected 2400 bursts in 6 years of observations, e.g. in 't Zand et al. 2004). This sample will be much less biased than previous ones, simply because the WFM is looking at every burster for 30% of the time, irrespective of the state of the burster. One example of the benefit that the WFM will provide is the fact that previous large samples often have just single burst detections for many sources. The resulting large and unbiased collection will enable much better studies of the thermonuclear burning behavior, and touch on questions like: how is the occurrence of superbursts related to the ordinary bursting behavior, how fast is the transition between unstable and stable burning and how does that relate to the accretion state, how does the long term bursting behavior depend on accretion composition (i.e., how does it compare between hydrogen-deficient ultracompact X-ray binaries and other LMXBs)?

Aside from shaping the light curves of individual bursts, nuclear reactions also strongly affect the overall burning behavior. For example, the mass accretion rate at which X-ray bursts are expected to disappear in favor of stable burning is greatly dependent on uncertain CNO-breakout reaction rates (e.g., Fisker et al. 2006; Keek et al. 2014). The WFM's large burst sample will map out the precise transitions between the burning regimes in different sources.

### 3.5 Synergy with laboratory experiments

It is a high priority of the low energy nuclear physics community to understand the nuclear reaction sequences in stellar hydrogen and helium driven explosions. X-ray bursts are the most common observable thermonuclear explosions, and sample the most extreme conditions of temperature and density of any directly observable, proton-rich environment. They are therefore the primary laboratory for such nuclear processes. The use of X-ray bursts as nuclear physics laboratories requires experimental data on the rates of the various nuclear reactions that may possibly occur in an X-ray burst, and comparison of burst simulations based on nuclear reaction networks with observations to validate models and to identify the actual reaction sequences.

The production of rare, unstable, isotopes in X-ray bursts is therefore a significant part of the nuclear astrophysics motivation for radioactive beam facilities around the world, including the 730 M\$ Facility for Rare Isotope Beams (FRIB), currently under construction at Michigan State University in East Lansing (USA), and the 1 B€ Facility for Antiproton and Ion Research (FAIR), under construction in Darmstadt (Germany). Major investments are being made in experimental equipment to measure astrophysical reaction rates in X-ray bursts.

However, a major drawback of using X-ray bursts as laboratories for the study of nuclear reactions has been the lack of specific observable signatures. While X-ray light curves provide some constraints, they are powered by the combined energy production of a large number of reactions and provide relatively weak constraints. A detection by *LOFT* of compositional signatures of the X-ray burst ashes, for example in the form of absorption edges (Weinberg et al. 2006), would dramatically strengthen observational constraints of nuclear processes in X-ray bursts and offer much improved validation pathways for X-ray burst models.

The nuclear physics community has a strong interest in using an improved understanding of nuclear reactions in X-ray bursts to address open astrophysical questions. With the construction of FRIB and other major equipment developments, significantly improved nuclear data can be expected over the next 10–20 years. A new X-ray mission, with advanced capabilities for the observation of X-ray bursts such as *LOFT*, is therefore particularly timely. The improved nuclear physics will significantly enhance the science reach. For example, superbursts will serve as more precise thermometers, and burst light curves and compositional signatures can be used to constrain much more precisely accretion rates, accreted composition, and the resulting composition of the neutron star crust. An advanced observational X-ray capability during the lifetime of FRIB will provide the opportunity to adjust X-ray burst models based on observations, and then tailor experiments to address new nuclear physics needs.

## 4 Non-uniform surface emission

### 4.1 Burst oscillations

About 20% of all bursts in the RXTE-PCA database exhibit oscillations (Galloway et al. 2008) that must be due to non-uniform emission patterns on the surface (see review by Watts 2012). They occur sometimes during burst rise and often during decay, measure 2–20% in rms amplitude and have a frequency that, in several cases, is confirmed to be that of the NS spin within a few Hz. The 11–620 Hz frequencies often show upward drifts of a few Hz throughout the burst tail. The oscillations during burst rise are most easily explained by a hotspot expanding from the point of ignition (Strohmayer et al. 1997). The oscillations during the tail could be caused by non-uniform emission during the cooling phase (e.g., because of different depths of fuel over the surface or cooling wakes; Spitkovsky et al. 2002; Zhang et al. 2013), but an interesting alternative involves global surface modes that may be excited by the motion of the deflagration front across the surface. This was first suggested by Heyl (2004), and further theoretical work was performed by (amongst others) Heyl (2005), Cumming (2005), Lee & Strohmayer (2005), Piro & Bildsten (2005) and Berkhout & Levin (2008).

The hotspot hypothesis is straightforward: ignition commences in a single location and spreads laterally over the surface. Numerical calculations show that speeds of deflagration fronts are expected to take $\sim 1$ s to engulf the entire neutron star (Fryxell & Woosley 1982; Cavecchi et al. 2013). The effects of flame spreading and the latitude of ignition are noticeable in the profile of the burst rise (Maurer & Watts 2008), but the evidence is still suggestive and inconclusive because of insufficient data. Only 2.5% of all RXTE-detected bursts show oscillations during the rise and only 1% show oscillations with long enough duration to allow investigation of their evolution, mostly at marginal significance, as shown by Chakraborty & Bhattacharyya (2014). This paper also shows that amplitudes in general decrease with time during burst rise, consistent with the flame spreading scenario, and that the profile of this decrease is consistent with a latitude-dependent flame spread speed, as would result from the Coriolis force (Spitkovsky et al. 2002; Bhattacharyya & Strohmayer 2007).

Surface modes, if they play a role in burst tail oscillations, are predicted to have a rich phenomenology (e.g., Piro & Bildsten 2005), including frequency drifts and multiple oscillations. No comprehensive model exists yet, but Heyl (2004) and Piro & Bildsten (2005) find one favorable mode for burst oscillations: the $l=2$, $m=1$ surface r-mode where $l$ and $m$ are the latitudinal and azimuthal quantum numbers. An important issue to be addressed for the surface r-mode model is that the predicted frequency drift is an order of magnitude larger ($\sim 10$ Hz) than the observed values (Watts 2012). Another more general issue is why burst oscillations are not seen in the majority of bursts, even among bursts from the same neutron star. This is not only an effect of sensitivity (e.g. Chakraborty & Bhattacharyya 2014). Other processes seem to influence the amplitudes of burst oscillations, such as the atmospheric composition, which influences the thermal heat transport time (Cumming & Bildsten 2000), and the ignition latitude (Maurer & Watts 2008). More sensitive measurements are needed to explore these effects.

Different ideas may be necessary to explain burst oscillations from LMXBs with accretion-powered pulsars. In these sources, the burst oscillation frequency can change sharply, by up to several Hz, in the rise, but there is little to no evolution of the frequency in the burst tail. The magnetic field may well play an important role in such sources; for example, Cavecchi et al. (2011) proposed that a dynamical magnetic flame stalling mechanism may be important.

With *LOFT*-LAD it will become possible to make great progress in understanding burst oscillations and employing them to study the physics of flame propagation and surface layer conditions. Burst oscillations can be measured at unprecedented detail, down to amplitudes of 0.6% rms (for a 1-s exposure and a brightness of 4 Crab), which will allow detection of shorter-lived signals, faster drifts and at lower burst fluxes (i.e., fainter bursts and burst rises and tails). By stacking bursts of the same source with similar burst profiles, the sensitivity may be improved to order 0.1% per second of the combined burst profile. This will allow for instance the following studies:

- Search for additional, weaker, oscillations in burst tails, including those with larger and faster frequency drifts as would be applicable to higher order r-modes (Piro & Bildsten 2005; Strohmayer & Mahmoodifar 2014). This will help determine what role, if any, surface r and g modes and their magnetic modifications (Heng & Spitkovsky 2009) play in burst tail oscillations, and will be possible on time scales as short as 10 s for single bursts and shorter for stacked bursts. Evidence for surface modes or hot spots may be derived from high-quality measurements of energy-dependent phase lags (Artigue et al. 2013).

- Obtain conclusive evidence that an expanding hotspot is responsible for burst oscillations during burst rises, by verifying for every oscillation that the fractional amplitude and phase evolve in a way consistent with hot spot expansion.

- Measure ignition latitudes for as many bursts as possible, and determine whether there are preferred latitudes (e.g., Spitkovsky et al. 2002; Cooper & Narayan 2007b). Determine how the Coriolis force is acting on flames by modeling the amplitude evolution as a function of NS spin frequency and ignition latitude.

- Search for (temporary) flame stalling (e.g., in multi-peaked bursts, see Bhattacharyya & Strohmayer

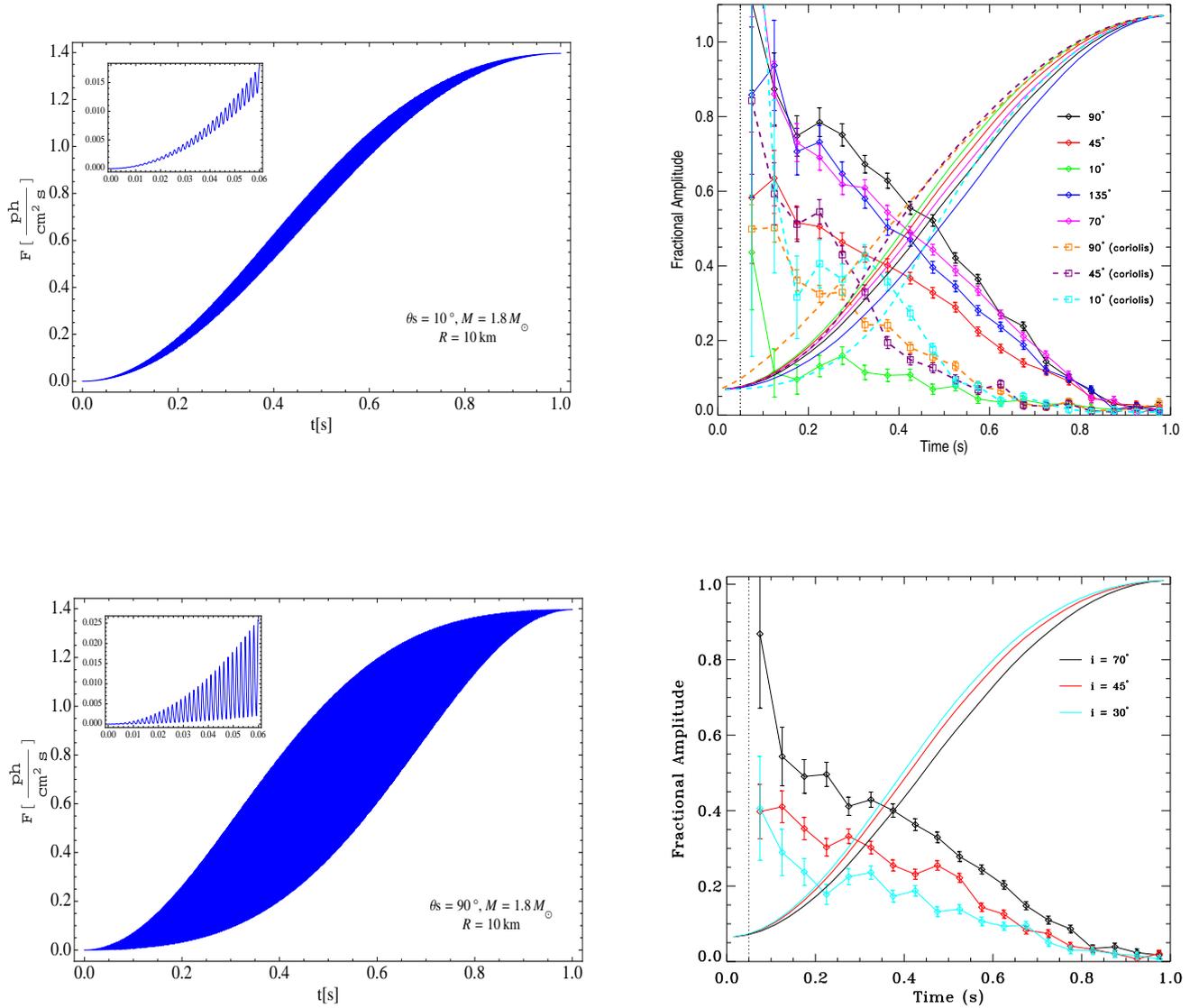

**Figure 2:** Simulations of flame spreading on the frequent and bright burster 4U 1636−536 (spin frequency 582 Hz). *(Left column:)* models of a signal from an expanding hotspot as it covers the NS in 1 s, for ignition polar angles (i.e., measured from the spin north pole) of $\theta = 10°$ (top) and $90°$(bottom). The angle between the NS spin axis and the line of sight is $i = 70°$. *(Right column:)* Simulated LAD measurements of the fractional amplitude after subtraction of background and non-burst flux, indicated by the circled data points with error bars connected with solid lines, for $i = 70°$ and a variety of polar angles $\theta$. The measurements connected with dashed lines are after including the Coriolis force on flame front propagation. The smooth solid and dashed lines running from bottom left to top right present the associated average model light curves in the same color coding as for the data points. The bottom plot shows fractional amplitude for $\theta = 45°$ and three inclination angles $i$.

2006a,b; Zhang et al. 2009), and determine which process is responsible for that (magnetic or equatorial; Cavecchi et al. 2011, 2014).

- Measure flame spreading in bursts with deep ignition (e.g., carbon-fueled superbursts). This could have an impact in understanding the dynamics of other thermonuclear events in astrophysics (e.g. classical novae or Type Ia SNe).

- Increase substantially the number of NS spin measurements in accreting NSs, allowing for a significantly better understanding of the spin distribution and possibly challenging the highest NS spin rates which can constrain the NS EOS, by detecting burst oscillations at lower amplitudes in more systems (see, e.g., Maccarone et al. 2015).

- In accreting ms pulsars, follow the spectral behavior during the transition from burst oscillations to pulsations as a means to find the origin of both types of oscillation.

We illustrate the potential of *LOFT* for the study of flame spreading by the simulations shown in Fig. 2. These simulations entail a constantly spreading disk-shaped hotspot emitting a black body spectrum of 2 keV, influenced by relativistic effects (Doppler shifts, relativistic aberration, gravitational redshift and light bending in a Schwarzschild geometry). We simulated LAD measurements of these oscillations for various ignition latitudes and an optimum observer inclination angle of 70°. A standard LAD background was assumed as well as an accretion flux of the source that is 7% of the peak burst flux, both of which were subtracted from the burst prior to amplitude determination. The amplitude was measured from folded pulse profiles per 0.05 s of integration. The fluxes and NS spin were assumed to be equal to that of 4U 1636−536, a source which is envisaged to be observed with *LOFT* for about 400 ks to satisfy core science goals. Three oscillations were simulated that include the possible effects of the Coriolis force to the extent calculated by Spitkovsky et al. (2002), see dashed curves in Fig. 2 (top right). The simulation clearly shows the richness of the data that can be obtained with *LOFT*. These kinds of measurements will enable the confirmation of the expanding hotspot model for burst oscillations during the rise, determination of ignition latitude for a number of bursts and measure the effect of the Coriolis force on flame propagation. The combination of the fractional amplitude evolution and the light curve will provide support to the hotspot model. Further diagnostic power lies in the spectral evolution of the burst signal, which is included in the modeling, but not illustrated in this simulation. A rotating hotspot will induce an oscillating Doppler shift of up to 10% in photon energy (for $v = 600$ Hz and $R = 10$ km) that has a phase offset of 0.2–0.25, the precise value of which is sensitive to relativistic effects, and should be detectable with LOFT.

Sometimes flames appear to propagate much faster than expected for a deflagration and burst oscillations are not present. The rise times to the Eddington limit in those cases are of order 1 ms. Very fast precursor events are involved, with durations of a few tens of milliseconds which are only detectable with the most sensitive instruments (in 't Zand & Weinberg 2010; in 't Zand et al. 2014). They indicate extreme explosive shell expansion with velocities up to tens of percent of the speed of light (in 't Zand et al. 2014) and the capability to disturb the accretion disk (in 't Zand et al. 2011). The LAD will be able to measure precursors in great detail, down to perhaps 10 microseconds. Superburst modeling predicts a different kind of precursor: shock breakout which lasts merely microseconds (Weinberg & Bildsten 2007; Keek & Heger 2011) and cannot be detected with any instrument except perhaps the LAD. The shock breakout may be a probe of the density profile in the neutron star ocean and atmosphere and of the ignition depth environment.

The RXTE-PCA measurements of the above-mentioned brightest and shortest precursor show rich and interesting non-periodic variability at (sub-)ms time scales which may be related to non-uniform emission patterns or structured winds or shells. The LAD will be able to pioneer this unexplored terrain in the field of km-sized structures.

## 4.2 Synergy with numerical calculations

Simulations of X-ray bursts are challenging because of the large range in length and timescales involved. The burning zone thickness is centimeters in scale, the scale height of the atmosphere is meters and the radius of the neutron star is ∼10 km. Current calculations cannot resolve all of these length scales simultaneously, so approximations are made. Current simulations of X-ray bursts largely fall into three categories. On the largest scale are global models of the spreading of a hot spot on a rotating neutron star using the shallow water equations (Spitkovsky et al. 2002). These use a simplified vertical structure that does not capture the details of the burning and rely on a parameterized flame speed, but demonstrate the importance of the Coriolis force in confining the burning. On the intermediate scale are the models of Cavecchi et al. (2013) which use wide-aspect ratio zones and a vertical hydrostatic equilibrium constraint to circumvent the restrictive Courant condition on timesteps. These calculations can model a propagating flame, but the wide-aspect ratio in the lateral direction means that they cannot fully resolve the burning zone or capture the effects of turbulence reliably. On the smallest scale are convection-in-a-box calculations, such as those of Malone et al. (2014). These can only resolve tens of meters laterally - too small to see the effects of rotation, but using a low Mach number approach, they can capture the details of the nucleosynthesis and accurately model the convection over many tens to hundreds of turnover times. Taken together, these three approaches provide a complementary picture of X-ray burst hydrodynamics. It is interesting to note that all three approaches use novel algorithms instead of the traditional fully-compressible hydrodynamics codes that dominate astrophysics.

Over the next 5 years we can expect these various approaches to continue to mature, and new algorithmic developments and faster supercomputers will allow the different simulation regimes to begin to better inform one another to build a complete picture of X-ray burst hydrodynamics and nucleosynthesis. On the small scales, algorithmic refinements may include acoustics and magnetic fields. The simulations here will push to larger domains and explore how the convection alters the appearance of the photosphere. These simulations can also provide the basis for a subgrid burning model that can be used by the larger scale simulations. On large scales, simulations will explore how the magnetic field affects hydrodynamics and flame propagation in cases where the former may play a significant role, as in accreting X-ray pulsars. Last but not least, general relativistic effects (time dilation and gravitational redshift) need to be included in multi-dimensional calculations.

A missing ingredient in all of these simulations is radiative transfer to connect the results directly to the observations. This deficiency should be addressed over the next few years as well, and may require leveraging the developments from the supernova community to do post-processing of simulation results.

## 5 Atmospheric conditions

Atmospheric conditions of accreting neutron stars have not yet been measured in much detail. Spectroscopy has proven hard because the accretion disk radiation dominates the signal outside bursts, and during bursts the signal is not present for long enough to collect the number of photons required for a sensitive spectral study at high resolution. *LOFT*-LAD will introduce an improved capability for such a study.

The atmospheric conditions of accreting neutron stars are set by the gravitational acceleration, composition and temperature. The composition depends on that of the atmosphere of the companion star and pollution by the nuclear burning just beneath. The latter is the most interesting for our purposes. The nuclear burning in neutron stars, while theoretically generally well understood, lacks detailed observational constraints. Models of CNO and $3\alpha$ burning explain X-ray bursts well to first order, and progress is being made with more detailed studies of burst profiles (e.g. Woosley et al. 2004; Heger et al. 2007b,a; Keek et al. 2012), but it would be a tremendous asset if the composition of nuclear ashes could be measured directly.

The principal emission from the photosphere is continuum radiation. Continuum spectra of bursting neutron stars are very close to diluted black body spectra, due to a strong interaction between photons and electrons (e.g., London et al. 1984, 1986; Madej 1991; Madej et al. 2004). The color temperatures corresponding to these spectra are larger than the effective temperatures by a factor $f_c = 1.4$ to 1.8. However, some deviations of the X-ray burst spectra from black body are predicted in the *LOFT* bandpass by modern model atmospheres, both at luminosities close to the Eddington limit as well as at relatively low luminosities $\lesssim 0.1 L_{Edd}$ for atmospheres with the (over-) solar iron abundance (Suleimanov et al. 2011, 2012). Spectra of the luminous model atmospheres are narrower than the corresponding black bodies, because they approach the Wien spectrum. On the other hand, at low luminosities iron is not completely ionized and the absorption edge of H-like iron may be observed. To date, the effective area of the available detectors allowed to study spectral deviations from the black body only during the unusually long stable bright phases of X-ray bursts (see, e.g., van Paradijs et al. 1990; Kuulkers et al. 2002a). The deviations observed in the spectrum of GX 17+2 during a luminous photospheric expansion phase are similar to those expected for high-luminosity neutron star atmospheres (Fig. 3-left). *LOFT* capabilities will allow us to investigate the evolution of this kind of deviation with luminosity during late burst stages even for short bursts. A comparison with the predictions of the atmosphere models will give us understanding of the upper layers of the radiating neutron stars as well as allow to constrain their basic properties.

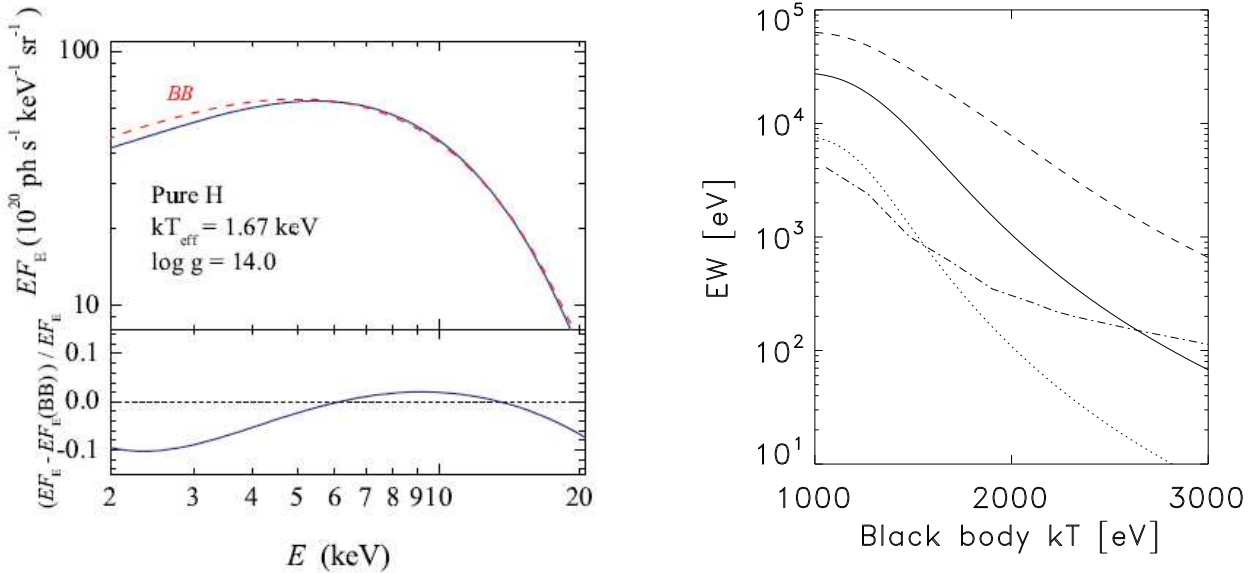

**Figure 3:** *(Left panel:)* The spectrum of the X-ray burst (in stellar frame) at a luminosity of $0.98L_{\mathrm{Edd}}$ (solid line) corresponding to the effective temperature of 1.67 keV and the best-fitting black body spectrum (with $T_{\mathrm{bb}} = 3.14$ keV and $f_c = 1.88$). The lower panel shows the ratio of the two spectra. *(Right panel:)* Theoretical predictions for the equivalent width of the Ni K edge at 8.33 keV for mass fractions of 0.01 (dotted curve), 0.1 (solid curve) and 1.0 (dashed curve) at the photosphere. For comparison, the $3\sigma$ detection limit is shown (dash-dotted curve) of a LAD measurement during a typical X-ray burst with peak flux 1 Crab on top of a persistent flux level of 0.1 Crab, plus background.

The prospects for direct detection of absorption features are promising, because the burning region is close to the photosphere (within a few meters), and during the most energetic events convection may dredge up the ashes into the photosphere, as some models suggest (Weinberg et al. 2006). This may imprint narrow features into the spectrum for part of the burst. There are tentative results of the direct detection of this with RXTE-PCA (in 't Zand & Weinberg 2010), but the 4 to 5 times improved spectral resolution, and 15 times improved photon collecting area of the LAD will provide much better measurements of those features. This may be possible in a few tens of energetic X-ray bursts with *LOFT*. If quick follow up of a superburst were feasible occasionally, large amounts of burst photons could be collected from a single burst and sensitivities for narrow spectral features would improve additionally by one to two orders of magnitude.

The question is in what way the ash dredge-up will affect the spectrum of the X-ray burst and whether *LOFT* will be able to detect it. Weinberg et al. (2006) predicts that absorption edges will appear in an otherwise unaffected thermal continuum spectrum, the edge energies and depths being governed by the population of ionization levels and abundances of heavy elements. The predictions are favorable towards a detection with the LAD (see Fig. 3-right). A step further is to model radiation transfer in the neutron star atmosphere, which might result in a change of the continuum as well. Preliminary results are shown in Fig. 4. The photospheric emission spectra were calculated with the radiative transfer code TLUSTY (Hubeny & Lanz 1995), assuming a gravitational field corresponding to a star of 1.4 $M_\odot$ and radius 10 km, and solar composition, but Fe a few times solar to model dredge up. The code treats the Fe atom fully in NLTE, i.e. the level populations are calculated explicitly from the rate equations, containing collisional and radiative terms. This is important since the spontaneous radiative decay rates between low lying levels and the ground state are extremely large, considerably exceeding the collisional rates. This leads to enhanced resonance scattering in the Ly-$\alpha$ line, and just like in the analogous situation in the hottest O stars, this can lead to Ly-$\alpha$ going from absorption to emission, at effective temperatures around and exceeding 2 keV. Likewise, the Lyman continuum appears in emission at high effective temperature. The details of the discrete spectrum will depend critically on the interaction between radiative transfer and the line-broadening mechanisms. Both effects can easily be detected with the LAD.

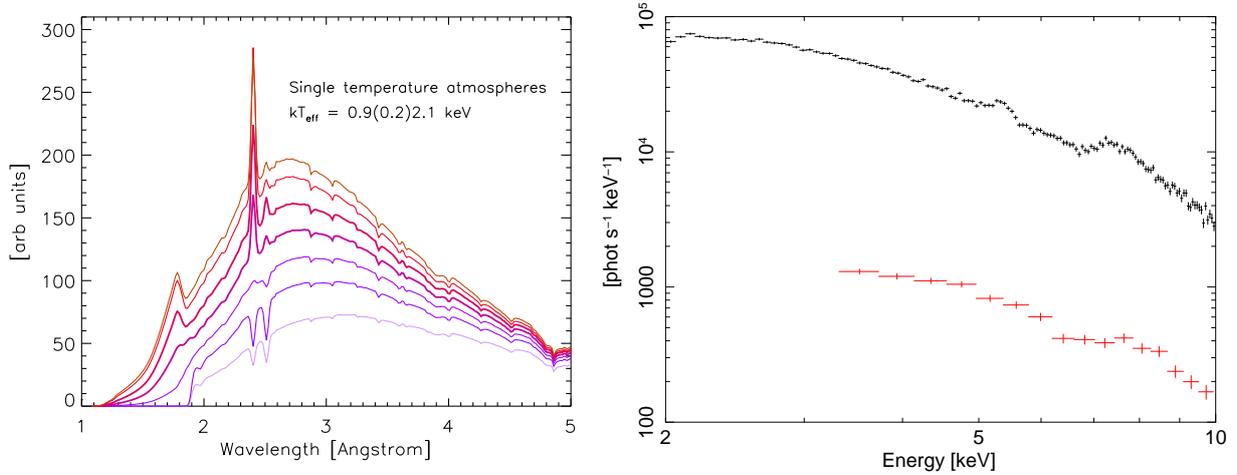

**Figure 4:** *(Left panel:)* TLUSTY models of NS photospheric spectra for various burst-like temperatures. *(Right panel:)* Simulation of a *LOFT*-LAD spectrum of a 1.9 keV TLUSTY spectrum at 1 Crab (exposure time 1 s; response employed for all events included). The lower set of data points refers, for comparison, to a simulation of a RXTE-PCA spectrum (3 active PCUs) with the same exposure time.

# 6 Accretion features

X-ray bursts emit high levels of X-ray radiation. In up to 20% of all bursts, they may also expel gas from the neutron star due to radiation pressure (20% is the percentage of bursts with photospheric radius expansion; Galloway et al. 2008). This radiation and gas will affect the accretion flow and its radiation. The burst radiation may, for instance, cool or heat coronal electrons due to Compton scattering. The radiation and gas may alter the geometry of the accretion disk. Part of the burst radiation may be reprocessed and reflected off the accretion flow into the line of sight. Thus, the emission that is observed during a burst actually consists of three contributions: direct burst emission, 'reflected' burst emission and accretion emission, all of which may vary considerably. To allow for a careful study of the burst emission alone, the second and third must be properly accounted for.

A few dramatic examples of burst-accretion interaction were detected through factor-of-two modulations of the burst flux during decay (e.g. Strohmayer & Brown 2002; in 't Zand et al. 2005, 2011; Degenaar et al. 2013). These bursts were rare superbursts or intermediate duration bursts with strong gas outflows, very likely in the form of winds and shells being expelled at high speeds from the neutron stars, and subsequent prolonged Eddington-limited states. Usually the interaction is more subtle, and correlated with the power of the nuclear reactions. In 't Zand et al. (2013) analyzed one Eddington-limited burst with Chandra and RXTE. Worpel et al. (2013) analyzed 332 such bursts with RXTE alone and both found that the non-burst emission during the burst became brighter by factors of up to several tens. This analysis was repeated on a larger selection of sub-Eddington bursts and again evidence was found that the non-burst emission increased during bursts, this time by factors of a few (Worpel et al., submitted). Models for this behavior vary. Worpel et al. (2013) attribute it to increased accretion due to the Poynting-Robertson (PR) effect (see also Walker 1992; Miller & Lamb 1993; Lamb & Miller 1995; Miller & Lamb 1996), while in 't Zand et al. (2013) argue it may be due to Compton scattering in the accretion disk corona. Resolution may come from *LOFT* measurements at high energies where the spectrum is dominated by accretion radiation which should always increase and decrease if the PR effect is active. The potential of such studies was recently shown on stacked RXTE burst data (Ji et al. 2013, 2014). LAD will do that on single bursts.

Figure 5: *LOFT*-LAD simulation of a burst spectrum consisting of a black body of $kT = 3$ keV including reflection against an ionized medium of ionization $\log \xi = 3$. The burst flux was chosen to be typical for common bright bursts of the nearby source 4U 1608−52 and the reflected component is 60% of the black body flux, similar to what is seen in the superburst from 4U 1636−536 (Keek et al. 2014) (Keek et al. 2015, in prep.). The top panel shows the measured LAD spectrum, the bottom panel the deviations with respect to a black body fit, in units of $1\sigma$ uncertainty per channel.

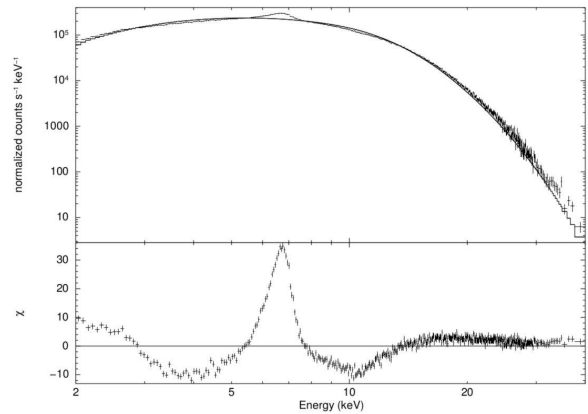

Reflection of burst radiation from an ionized medium has been detected in two superbursts with RXTE, through its hallmark iron-K emission line and absorption edge (Ballantyne & Strohmayer 2004; Keek et al. 2014). For one of the superbursts the location of the reflector could be constrained at 64 s time resolution. LAD will be able to perform this measurement for many short X-ray bursts at a higher time resolution and with smaller uncertainties, see Fig. 5. This will allow for the study of the evolution of accretion disks while irradiated by X-ray bursts on a much larger scale. It will provide a powerful probe into accretion physics, enabling measurement of for instance the ionization state and geometry of the accretion flow and potentially the viscous processes at work in the flow (Ballantyne & Everett 2005). Moreover, this will enable better separation of the neutron star emission from reflection and persistent components. Presently, reflection is not taken into account in such studies, nor is the evolution of the persistent flux, which introduces substantial systematic uncertainties in $M/R$ measurements (Poutanen et al. 2014; Kajava et al. 2014).

## 7 Conclusion

With respect to the previous workhorse in X-ray burst research, RXTE-PCA, *LOFT* will be a dramatic improvement in three areas. The LAD will collect roughly 15 times more photons per burst (compared to 5 active proportional counter units of the PCA; with respect to the RXTE mission-average of 3, this factor is even 25) and it will have a 4 to 5 times better spectral resolution enabling more accurate studies of absorption edges. The WFM will have an unprecedented high duty cycle for similarly sensitive instruments, resulting in the detection of tens of thousands of X-ray bursts, which is more than have been detected so far in the history of X-ray astronomy. This opens up the possibility to probe deeper into the neutron star and discover new rare phenomena in X-ray bursts. In Table 1, we summarize what *LOFT* can do for X-ray burst research and specify for each item how many X-ray bursts can possibly be studied in a 3-yr mission.

Table 1: List of new measurements that *LOFT* LAD and WFM can perform on X-ray bursts.

| Instr. | Burst feature | # bursts | Kind of measurement and goal |
|---|---|---|---|
| LAD | Oscillations in ordinary bursts | Hundreds | - Find fainter oscillations to increase burster population with oscillations |
| | | | - Find fainter oscillations, for instance in tail, to enlarge parameter space of oscillations |
| | | | - Study pulse profile with time to study flame spreading |
| | | | - Measure frequency drifts better to constrain models for oscillations in tails |
| | Oscillations in superbursts | Few | Study pulse profile with time to study flame spreading for deeper ignitions |
| | Continuum spectra | Hundreds | Measure peak flux more accurately by constraining better bolometric correction, study deviations from black body spectra |
| | Absorption edges | Tens | Obtain unambiguous detections, to constrain ash composition and gravitational redshift |
| | Time profiles | Hundreds | - Extend detection to lower fluxes, to probe cool tail phase (heating/cooling of deep layers) |
| | | | - Detect deviations inconsistent with cooling, to probe disturbances of accretion flow |
| | Low-luminosity bursts | Few | - Probe incomplete latitudinal burning |
| | | | - Probe unstable hydrogen CNO burning (pure H bursts) |
| | Superburst aftermath | Few | Obtain for the first time complete coverage of this phase, to probe marginal burning / burst quenching |
| | Precursors in ordinary bursts | Tens | Probe shell / accretion disk dynamics, fastest ignition regimes and flame spreading |
| | Precursors in superbursts | Very few | Detect them at shorter durations (sub-ms), to probe shock breakout and premature ignition of ordinary bursts |
| | Reflection features | Tens | Track disk ionization, surface density and accretion rate |
| WFM | Superbursts | Tens | - Constrain recurrence times |
| | | | - Measure accurate fluences by more complete coverage |
| | | | - Obtain more data on superburst onsets |
| | Low $\dot{M}$ bursters | Tens | Constrain population characteristics, probe low $\dot{M}$ accretion disk physics, probe low $\dot{M}$ burst regime |

In this paper we have not discussed $M/R$ measurements of NSs in detail. This is dealt with by the *LOFT* core science case (and derives, amongst other methods, from burst oscillations; Feroci et al. 2014, and the *LOFT* Yellow Book), but it is clear that further $M$ and $R$ constraints can come from the detection of gravitationally red-shifted narrow spectral features and light curves. Particularly, Eddington-limited X-ray bursts with photospheric expansion could be constraining in this respect, because GR effects should then change during the bursts in a well-defined manner.

We note that the ESA-L2 mission Athena is no match to the capabilities of the LAD for X-ray burst research, because its effective area, peaking at 1 keV at a value of $2\,\mathrm{m}^2$, decreases to about $(5-8) \times 10^3\,\mathrm{cm}^2$ for photon energies where the burst spectrum peaks during the brightest stages (2–5 keV). At this energy, Athena performs similar to RXTE-PCA (albeit with better spectral resolution), but is much smaller than the $(4-8) \times 10^4\,\mathrm{cm}^2$ for the LAD. At higher energies, this comparison is even less favorable for Athena.

## Acknowledgments

We sincerely thank E. Kuulkers, H. Worpel, E. Bozzo, A. Santangelo and J. Wilms for useful discussions, and E. Bozzo for help with Fig. 1. DA acknowledges support from the Royal Society, ALW and YC from NWO Vrije Competitie grant 614.001.201 (PI Watts), JC from ESA (Prodex nr. 90057), AR by Polish NSC grant Nr. 2013/10/M/ST9/00729, AH by an ARC Future Fellowship (FT120100363), DRB and LK from NASA ADAP grant NNX13AI47G and MZ from US DOE NP DE-FG02-87ER40317.

## References

Altamirano D., van der Klis M., Wijnands R., Cumming A., 2008, ApJ 673, L35

Artigue R., Barret D., Lamb F.K., et al., 2013, MNRAS 433, L64

Ballantyne D.R., Everett J.E., 2005, ApJ 626, 364

Ballantyne D.R., Strohmayer T.E., 2004, ApJ 602, L105

Berkhout R.G., Levin Y., 2008, MNRAS 385, 1029

Bhattacharyya S., Strohmayer T.E., 2006a, ApJ 636, L121

Bhattacharyya S., Strohmayer T.E., 2006b, ApJ 641, L53

Bhattacharyya S., Strohmayer T.E., 2007, ApJ 666, L85

Bildsten L., 1998, In: Buccheri R., van Paradijs J., Alpar A. (eds.) NATO Advanced Science Institutes (ASI) Series C, Vol. 515. NATO Advanced Science Institutes (ASI) Series C, p. 419

Cavecchi Y., Patruno A., Haskell B., et al., 2011, ApJ 740, L8

Cavecchi Y., Watts A.L., Braithwaite J., Levin Y., 2013, MNRAS 434, 3526

Cavecchi Y., Watts A.L., Levin Y., Braithwaite J., 2014, MNRAS, in press

Chakraborty M., Bhattacharyya S., 2014, ApJ 792, 4

Cooper R.L., Narayan R., 2007a, ApJ 661, 468

Cooper R.L., Narayan R., 2007b, ApJ 657, L29

Cornelisse R., Heise J., Kuulkers E., et al., 2000, A&A 357, L21

Cornelisse R., in 't Zand J.J.M., Verbunt F., et al., 2003a, A&A 405, 1033

Cornelisse R., in 't Zand J.J.M., Verbunt F., et al., 2003b, A&A 405, 1033

Cumming A., 2005, ApJ 630, 441

Cumming A., Bildsten L., 2000, ApJ 544, 453

Cumming A., Bildsten L., 2001, ApJ 559, L127

Cumming A., Macbeth J., 2004, ApJ 603, L37

Cumming A., Macbeth J., in 't Zand J.J.M., Page D., 2006, ApJ 646, 429

Degenaar N., Miller J.M., Wijnands R., et al., 2013, ApJ 767, L37

Feroci M., den Herder J.W., Bozzo E., et al., 2014, In: Society of Photo-Optical Instrumentation Engineers (SPIE) Conference Series, Vol. 9144. Society of Photo-Optical Instrumentation Engineers (SPIE) Conference Series, p. 2

Fisker J.L., Schatz H., Thielemann F.K., 2008, ApJ Supp. 174, 261

Fryxell B.A., Woosley S.E., 1982, ApJ 261, 332

Fujimoto M.Y., Hanawa T., Miyaji S., 1981, ApJ 247, 267

Galloway D.K., Muno M.P., Hartman J.M., et al., 2008, ApJ Supp. 179, 360

Heger A., Cumming A., Galloway D.K., Woosley S.E., 2007a, ApJ 671, L141

Heger A., Cumming A., Woosley S.E., 2007b, ApJ 665, 1311

Heng K., Spitkovsky A., 2009, ApJ 703, 1819

Heyl J.S., 2004, ApJ 600, 939

Heyl J.S., 2005, MNRAS 361, 504

Hubeny I., Lanz T., 1995, ApJ 439, 875

in 't Zand J., Verbunt F., Heise J., et al., 2004, Nuclear Physics B Proceedings Supplements 132, 486

in 't Zand J.J.M., Cumming A., van der Sluys M.V., et al., 2005, A&A 441, 675

in 't Zand J.J.M., Galloway D.K., Ballantyne D.R., 2011, A&A 525, A111

in 't Zand J.J.M., Galloway D.K., Marshall H.L., et al., 2013, A&A 553, A83

in 't Zand J.J.M., Jonker P.G., Markwardt C.B., 2007, A&A 465, 953

in 't Zand J.J.M., Keek L., Cavecchi Y., 2014, A&A 568, A69

in 't Zand J.J.M., Kuulkers E., Verbunt F., et al., 2003, A&A 411, L487

in 't Zand J.J.M., Weinberg N.N., 2010, A&A 520, A81

Jahoda K., Markwardt C.B., Radeva Y., et al., 2006, ApJ Supp. 163, 401

Ji L., Zhang S., Chen Y., et al., 2013, MNRAS 432, 2773

Ji L., Zhang S., Chen Y., et al., 2014, ApJ 782, 40

Kajava J.J.E., Nättilä J., Latvala O.M., et al., 2014, MNRAS 445, 4218

Keek L., Ballantyne D.R., Kuulkers E., Strohmayer T.E., 2014, ApJ 797, L23

Keek L., Heger A., 2011, ApJ 743, 189

Keek L., Heger A., in 't Zand J.J.M., 2012, ApJ 752, 150

Keek L., in 't Zand J.J.M., Kuulkers E., et al., 2008, A&A 479, 177

Kuulkers E., Homan J., van der Klis M., et al., 2002a, A&A 382, 947

Kuulkers E., in 't Zand J.J.M., Atteia J.L., et al., 2010, A&A 514, A65

Kuulkers E., in 't Zand J.J.M., van Kerkwijk M.H., et al., 2002b, A&A 382, 503

Lamb F.K., Miller M.C., 1995, ApJ 439, 828

Lasota J.P., 2001, New Astron. 45, 449

Lee U., Strohmayer T.E., 2005, MNRAS 361, 659

Lewin W.H.G., van Paradijs J., Taam R.E., 1993, Space Sci. Rev. 62, 223

Linares M., Altamirano D., Chakrabarty D., et al., 2012, ApJ 748, 82

London R.A., Howard W.M., Taam R.E., 1984, ApJ 287, L27

London R.A., Taam R.E., Howard W.M., 1986, ApJ 306, 170

Maccarone T., Wijnands R., Degenaar N., et al., 2015, In: LOFT White Paper., p. 1

Madej J., 1991, ApJ 376, 161

Madej J., Joss P.C., Różańska A., 2004, ApJ 602, 904

Malone C.M., Zingale M., Nonaka A., et al., 2014, ApJ 788, 115

Maurer I., Watts A.L., 2008, MNRAS 383, 387

Medin Z., Cumming A., 2011, ApJ 730, 97

Miller M.C., Lamb F.K., 1993, ApJ 413, L43

Miller M.C., Lamb F.K., 1996, ApJ 470, 1033

Paul B., LAXPC Team 2009, In: Kawai N., Mihara T., Kohama M., Suzuki M. (eds.) Astrophysics with All-Sky X-Ray Observations., p. 362

Peng F., Brown E.F., Truran J.W., 2007, ApJ 654, 1022

Piro A.L., Bildsten L., 2005, ApJ 629, 438

Poutanen J., Nättilä J., Kajava J.J.E., et al., 2014, MNRAS 442, 3777

Revnivtsev M., Churazov E., Gilfanov M., Sunyaev R., 2001, A&A 372, 138

Schatz H., Bildsten L., Cumming A., 2003, ApJ 583, L87

Schatz H., Gupta S., Möller P., et al., 2014, Nature 505, 62

Spitkovsky A., Levin Y., Ushomirsky G., 2002, ApJ 566, 1018

Strohmayer T., Bildsten L., 2006, New views of thermonuclear bursts. In: Lewin W.H.G., van der Klis M. (eds.) Compact stellar X-ray sources. Cambridge Univ. Press, Cambridge, p.113

Strohmayer T., Mahmoodifar S., 2014, ApJ 793, L38

Strohmayer T.E., Brown E.F., 2002, ApJ 566, 1045

Strohmayer T.E., Zhang W., Swank J.H., 1997, ApJ 487, L77

Suleimanov V., Poutanen J., Werner K., 2011, A&A 527, A139

Suleimanov V., Poutanen J., Werner K., 2012, A&A 545, A120

van Paradijs J., Dotani T., Tanaka Y., Tsuru T., 1990, PASJ 42, 633

van Paradijs J., Penninx W., Lewin W.H.G., 1988, MNRAS 233, 437

Walker M.A., 1992, ApJ 385, 661

Wallace R.K., Woosley S.E., 1981, ApJ Supp. 45, 389

Watts A.L., 2012, ARAA 50, 609

Weinberg N.N., Bildsten L., 2007, ApJ 670, 1291

Weinberg N.N., Bildsten L., Schatz H., 2006, ApJ 639, 1018

Woosley S.E., Heger A., Cumming A., et al., 2004, ApJ Supp. 151, 75

Worpel H., Galloway D.K., Price D.J., 2013, ApJ 772, 94

Zamfir M., Cumming A., Niquette C., 2014, MNRAS 445, 3278

Zhang G., Méndez M., Altamirano D., et al., 2009, MNRAS 398, 368

Zhang G., Méndez M., Belloni T.M., Homan J., 2013, MNRAS 436, 2276